\documentclass[aps,pra]{revtex4}
\usepackage{amsmath}
\usepackage{graphicx}
\usepackage{epsfig}

\newcommand{\be}{\begin{equation}}
\newcommand{\ee}{\end{equation}}

\def \be{\begin{equation}}
\def \ee{\end{equation}}
\def \ba{\begin{array}}
\def \ea{\end{array}}
\def \beq{\begin{eqnarray}}
\def \eeq{\end{eqnarray}}

\def \etal{{\it {et al}}}

\def \a{{\alpha}}

\def \nd{{^{\vphantom{\dagger}}}}
\def \yd{^\dagger}
\def \av#1{{\langle#1\rangle}}

\begin{document}

\title{How to study correlation functions in fluctuating
Bose liquids using interference experiments}


\author{Vladimir Gritsev$^1$, Ehud Altman$^2$, Anatoli Polkovnikov$^3$, Eugene Demler$^1$}{
\affiliation{$^1$Department of Physics, Harvard University,
Cambridge, MA 02138, USA\\ $^2$Department of Condensed Matter
Physics, The Weizmann Institute of Science Rehovot, 76100, Israel,
$^3$Department of Physics, Boston University, Boston, MA 02215, USA}

\begin{abstract}

Interference experiments with independent condensates provide a
powerful tool for analyzing correlation functions. Scaling of the
average fringe contrast with the system size is determined by the
two-point correlation function and can be used to study the Luttinger
liquid liquid behavior in one-dimensional systems and to observe the
Kosterlitz-Thouless transition in two-dimensional quasicondensates.
Additionally, higher moments of the fringe contrast can be used to
determine the higher order correlation functions. In this article we
focus on interference experiments with one-dimensional Bose liquids
and show that methods of conformal field theory can be applied to
calculate the full quantum distribution function of the fringe
contrast.

\end{abstract}

\maketitle

Correlation functions provide a convenient language for characterizing
quantum states. First-order coherence in optics, which
underlies classical Young's interference experiments, is mathematically
equivalent to the factorizability of the first order correlation
function $G^{(1)}(x_1,x_2)= \langle E^{(-)}(x_1) E^{(+)} (x_2)
\rangle$, where $E^{(\pm)}$ are the electric field components which
vary as $e^{\pm i \omega t}$.  Photon bunching in Hanbury-Brown and
Twiss experiments can be understood using the second order correlation
function $G^{(2)}(\tau)= \langle E^{(-)}(t) E^{(-)}(t+\tau)
E^{(+)}(t+\tau) E^{(+)}(t) \rangle$.  The same correlation function
describes photon antibunching in quantum mechanical states of light
such as number states of photons\cite{walls}. Analogously, in
condensed matter physics many common experiments can be understood as
probes of the appropriate two point correlation functions. For example,
conductivity measurements (dc or finite frequency) correspond to
current-current correlation functions \cite{mahan}, angle resolved
photoemission\cite{jccampuzano,zxshen} and tunneling
experiments\cite{wolf} probe single particle Green's functions, X-Ray
and neutron scattering experiments can be related to charge and spin
correlation functions\cite{ashcroft}.

An important question for current experiments with ultracold atoms is
finding new methods of characterizing strongly correlated many-body
states, such as systems near Feshbach resonances, rotating condensates,
atoms in optical lattices, low-dimensional systems (see
ref. \cite{anglin} for a review). The existing experimental toolbox
includes Bragg scattering, which measures the dynamic
structure factor \cite{ketterle2000}
(i.e. the imaginary part of the density-density correlation function),
 and the RF spectroscopy (see e.g.
Refs. \cite{chin2004,storfele2006}), which is essentially
equivalent to finite frequency conductivity measurements in solid
state systems. These techniques measure different types of two point
correlation functions. An interesting question to ask is whether one
can do experiments with systems of cold atoms that would measure
higher order correlation functions. In two recent papers
\cite{pad,interference2} we argued that this can be done using
interference experiments with independent condensates (see also Ref.
\cite{niu2006}). In this article we will review these ideas while
focusing on a specific case of one-dimensional condensates.

We mention in passing that another interesting technique, which is
unique to systems of cold atoms, is the time of flight experiments (see
e.g. Ref. \cite{pitaevski}). When expansion of atoms released from the
trap is ballistic, the momentum distribution of atoms inside the
original system gets mapped into the density distribution after the
expansion. Measurements of quantum noise after the expansion have been
used to study Hanbury-Brown-Twiss correlations for bosons and
fermions in optical
lattices\cite{foelling,nist,mainz_fermions} and pairing correlations
on the molecular side of the Feshbach resonance\cite{greiner}. The
correlation function measured in such experiments is a non-local
operator from the point of view of the original system but should
provide ``smoking gun'' signatures of several many-body phases
including antiferromagnetically ordered Mott states and paired states
of fermions\cite{altman}.

The interference experiments we consider are shown in
Fig.~\ref{fig:tubes}.
\begin{figure}[ht]
\includegraphics[width=8.0cm]{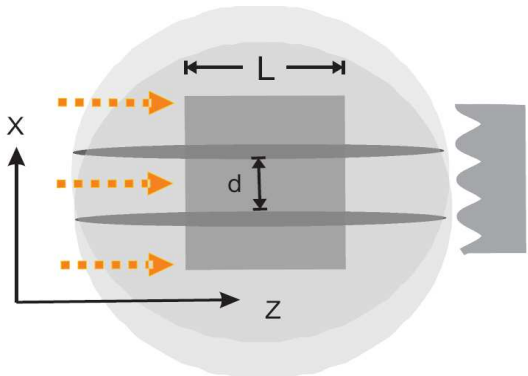}
\caption{Experimental setup}
\label{fig:tubes}
\end{figure}
Two independent quasi-condensates are allowed to expand in the
transverse direction. After sufficient expansion, the atom
distribution is measured using the imaging beam.  Everywhere in this
paper we consider the two condensates to be identical, although our
analysis can be generalized to the case of different condensates.  In
the set-up we discuss, the measuring beam is oriented along the axis
of the original cloud. This beam integrates over local interference
patterns and the fringes that we see on a detector are a result of
averaging over the entire imaging length.  In the presence of thermal
or quantum fluctuations, interference patterns are not in phase at
different points and the resulting interference fringes have a reduced
contrast. Such reduction carries information about fluctuations in the
original clouds. The main topic of this article is a non-trivial
information that we can extract by analyzing such reduced contrast. We
point out that an alternative set-up, which is also possible with
current experiments, is to measure not along the system axis but from
the side. This provides a local picture of interference fringes rather
than the integrated one. A convenient way to analyze such data would
be to integrate interference patterns numerically and study how the
average contrast changes with the system size (see discussion
below). So the main advantage of the latter set-up is a dramatic
reduction in the number of measurements needed to determine the
scaling of interference fringes. Conceptually, however,  analysis is
identical to the case that we discuss in this paper.  We assume that
before the expansion, atoms are confined to the lowest transverse
channels of their respective traps and that the optical imaging length
$L$ (which is smaller than the size of the system in the axial
direction) is much larger than the coherence length of the
condensates. This allows us to use an effective Luttinger liquid
description of the interacting bosons\cite{cazalilla}.

The quantum observable
corresponding to the interference amplitude of the two
condensates~\cite{pad} is given by
$A_Q=\int_0^{L} dz a\yd_1(z)a\nd_2(z)$.
Here $a_1$ and $a_2$ are the bosonic operators in the two systems
before the expansion, and the integrals are taken along the
condensates. $L$ is the imaging length that is in general smaller than the full
condensate length.  If the two condensates are
decoupled from each other, the expectation value of $\langle A_Q
\rangle $ vanishes. This does not mean that $|A_{Q}| $ is zero in
each individual measurement but rather shows that the phase of $A_{Q}$
is random. Hence the position of interference fringes is completely
random from shot to shot. The quantum mechanical expectation value is
defined as a result of averaging over many experimental
runs. Superimposed interference patterns with random phases wash each
other out and interference fringes averaged over many shots
disappear\cite{hadzibabic01}.  However, what we are interested in is
the amplitude of the fringes in an individual measurement.  Thus we
should consider an observable that does not involve the random phase
of $A_Q$. We take
\be \!\!\!|A_{Q}|^2=\!\!\int_0^L\!\!\int_0^L\!\!dz_1 dz_2\,
a_1^\dagger(z_1) a_1(z_2)\,
a_2^\dagger(z_2) a_2(z_1)\,
\label{Ap2}
\ee
To simplify calculations we consider the limit
of a large system when we are allowed to take the normal ordered
expression\cite{ap2006}.
Taking the expectation value of the last equation produces a nontrivial
result since on the right hand side operators that correspond to different
clouds decouple.
Assuming that the condensates are identical  gives
\be
\langle |A_{Q}|^2 \rangle = L\int_0^L dz \av{a\yd(z)a\nd(0)}^2.
\label{main}
\ee
Thus the average intensity of the interference pattern depends on the
two point correlation function along the individual one-dimensional
condensates.  In this
regime the long distance correlations decay as a power law within
the imaging length $L$, $\langle
a^\dagger(z) a(0)\rangle\sim z^{-1/2K}$,
where $K$ is the Luttinger parameter. For bosons with a repulsive
short-range potential, $K$ ranges between $1$ and $\infty$, with $K=1$
corresponding to strong interactions, or ``impenetrable'' bosons, while
$K\to \infty$ for non-interacting bosons. Using the power law
correlations in (\ref{main}),  we find that the interference intensity
scales as a non-trivial power of the imaging length $\langle |A_Q|^2
\rangle= L^{2-1/K}$.  This property can be used as a sensitive probe
of the Luttinger parameter. In the non-interacting limit
($K\to\infty$), the scaling is linear, $\bar{A}_Q \equiv \sqrt{\langle
|A_Q|^2 \rangle}
\sim L$, as expected for
a fully coherent system. Interestingly, in the impenetrable limit,
$\bar{A}_Q\sim \sqrt{L}$, as for short range exponentially decaying
correlations. One may ask whether it makes sense to discuss zero
temperature limit when experiments are always done at finite
temperature. It is important to realize that experiments are done in
systems of finite size.  At finite temperatures and at long
distances, correlations always decay exponentially. However, long
distances means distances larger than the correlation length $\xi_T
\sim {\rm v}_{\rm s}/T$, where $ {\rm v}_{\rm s}$ is the sound velocity.
As the temperature is lowered, $\xi_T$ increases and at some point
becomes larger than the system size. At this point (and at all lower
temperatures), correlations within the system size are those of the
zero temperature system. It may be useful to make one remark regarding
experimental constrains for observing quantum Luttinger liquid
behavior with ultracold atoms. A common parameter used to characterize
the strength of interactions between particles is the ratio of the
interaction energy to the kinetic energy $\gamma = m g/{\hbar}^2 n$.
Here $m$ is the mass of atoms, $g$ is the interaction strength, and
$n$ is the one-dimensional density. To observe strongly interacting
Bose liquid we want $\gamma$ to be large, which may be achieved by
reducing the density.  However we also need to satisfy the condition
$\xi_T\gg L$. If we use Bogoliubov expression for ${\rm v}_{\rm s}$ we
find $L^{-1}
\sqrt{ng/m} >> T$.  Hence for a smaller density it takes a lower
temperature to reach the quantum limit. In experiments one
needs to find the balance between having the interesting regime of
strong interactions and reaching the quantum limit.

We point out that analysis presented above for one dimensional systems
can be easily generalized to interference between planar condensates
at finite temperatures\cite{pad}. The power law decay of correlations,
and thus the scaling of the interference intensity with imaging size
is related in this case to the superfluid stiffness of the
condensates. This was used by Hadzibabic {\it \etal} \cite{hadzibabic}
to observe the universal jump of superfluid stiffness across the
Kosterlitz-Thouless transition in pancake condensates.

So far we have discussed the average intensity of the interference pattern
and showed that it contains information about two-point correlation
functions. It is important to realize that interference
experiments correspond to a classical measurement of a quantum
mechanical state. Hence they contain an intrinsic quantum mechanical
noise and the result of each individual measurement will be different
from the average value. Generalization of the argument
that led to eq. (\ref{main}) shows that higher moments
of the distribution function of interference amplitudes correspond to
high-order correlation functions. Hence the knowledge of the entire
distribution function reveals global properties of the system that
depend on very high order non-local correlation functions.

 From equation (\ref{Ap2}) one finds that for two one-dimensional
condensates higher moments of the interference fringe amplitude are
given by~\cite{pad}
\begin{eqnarray}
\langle |A_{Q}|^{2n} \rangle = A_0^{2n} Z_{2n},\;
\mbox{where}\;A_0=\sqrt{C\rho\xi_h^{1/K} L^{1-1/K}},
\label{HigherMoments}
\end{eqnarray}
where $C$ is a constant of order unity,
$\rho$ is the particle density in each condensate, $\xi_h$ is the
short range cutoff equal to the healing length. The coefficients
$Z_{2n}$ in Eq.~(\ref{HigherMoments}) are given by\cite{note}:
\beq
Z_{2n}(K)=\int_0^{2\pi}\!\dots\!\int_0^{2\pi}\prod_{i=1}^{n}\frac{du_{i}}
{2\pi}\frac{dv_i}{2\pi}
\left|{\prod_{i<j} 2\sin\left({u_i-u_j\over 2}\right)\prod_{k<l}
2\sin\left({v_k -v_l\over 2}\right)\over
\prod_{i,k}2\sin\left({u_i-v_k\over 2}\right)}\right|^{1/K},
\label{z2n}
\eeq

The coefficients $Z_{2n}$ originally appeared in the grand canonical
partition function of a neutral two-component Coulomb gas on a
circle
\be
Z(K,x)=\sum_{n=0}^{\infty} {x^{2n}\over (n!)^2} Z_{2n}(K).
\label{z}
\ee Here $x$ is the fugacity of Coulomb charges and $Z_{2n}$ describes
contributions from configurations with $2n$ charges (i.e.  canonical
partition functions). The partition function (\ref{z}) with $K>1$
describes several problems in statistical
physics (see Ref \cite{saleur} and references therein). It is
also related to the problems of an impurity in a Luttinger
liquid \cite{kanefisher,interference2} and Kondo effect\cite{FLS2}.

When describing interference experiments it is convenient to define
the normalized amplitude of interference fringes
$\a=|A_{Q}|^{2}/A_{0}^{2}$.  From (\ref{HigherMoments}) we find that
$\langle
\a^{2n} \rangle = Z_{2n}$, so by performing experiments that
measure the distribution function $W(\a)$, we get direct access to
the partition function (\ref{z}). We point out that $W(\a)$ can be
used to compute all moments of $|A_{Q}|^{2}$, and therefore
contains information about high order correlation functions of the
interacting Bose liquids.

From the Taylor expansion of the modified Bessel function we find
\be
Z(K,x)=\int_0^\infty W(\a)\, I_0(2x \sqrt{\a})\, d \a.
\label{z1}
\ee
Inverting Eq.~(\ref{z1}) allows one to express the probability $W(\a)$
through the partition function $Z(K,x)$. Noting that
$I_0(ix)=J_0(x)$ and using the completeness relation for Bessel
functions, $\int_0^\infty J_0(\lambda x) J_0(\lambda y) |x|\lambda
d\lambda=\delta(|x|-|y|)$, we obtain
\be
W(\a)=2\int_0^\infty Z(K,i x)J_0(2x \sqrt{\a})x dx.
\label{wz}
\ee

The remaining problem is to calculate the partition function
$Z(K,ix)$.  In this
paper we will use a  method based on the studies
of the integrable structure of conformal field
theories~\cite{BLZ1-3}. In particular, it was shown that the vacuum
expectation value of Baxter's ${Q}$ operator (central to the
integrable structure of the models\cite{baxter}), coincides with the grand
partition function of interest:
\be\label{al}
Q^{\rm vac}(\lambda)= Z(K,-i x),
\ee
where $x$ is related to the spectral parameter $\lambda$,
$x=\pi\lambda/ \sin(\pi/2K)$.
It was conjectured in Refs.~\cite{DT,BLZ}
that the vacuum expectation value $Q^{\rm vac}(\lambda)$
is proportional to the spectral determinant of the single
particle Schr\"{o}dinger
equation
\begin{equation}
-\partial_{x}^{2}\Psi(x)+\left(x^{4K-2}-\frac{1}{4x^{2}}\right)\Psi(x)=E\Psi(x),
\label{schrod}
\end{equation}
So, $Q^{\rm vac}(\lambda)=  D(\rho\lambda^{2})$, where $\rho
=(4K)^{2-1/K}[\Gamma(1-1/(2K))]^{2}$, $D(E)$ is the spectral
determinant defined as $D(E)=\prod_{n=1}^{\infty}(1-E/E_n)$, and
$E_n$ are the eigenvalues of (\ref{schrod}). Thus, we have
\begin{equation}
Z(K,ix)=\prod_{n=1}^{\infty}\left(1-\frac{\rho\lambda^2}{E_{n}}\right).
\label{conj}
\end{equation}

 The distribution function $W(\a)$ is shown in Fig. \ref{figK} for
several values of $K$. For $K$ close to 1 (Tonks-Girardeau limit),
$W(\a)$ is a wide Poissonian function, which gradually narrows as $K$
increases, finally becoming a narrow $\delta$-function at $K\to\infty$
(limit of noninteracting bosons). Note that the distribution function
remains {\it asymmetric} for arbitrarily large $K$. In fact we find
that $W(\tilde\a-1)$, where
$\tilde{\alpha}=\a/\av{\a}=|A|^2/\av{|A|^2}$, tends to a universal
scaling form, parameterized by a single number characterizing the
width of the distribution:
$\delta\equiv\sqrt{\langle\tilde\a^2\rangle-1}\approx \pi/\sqrt{6}K$.
We conjecture that the limiting form of
$W(\tilde\a)$ is the Gumbel distribution function
\cite{interference2}. The appearance of the Gumbel distribution in
this problem is not surprising. This distribution was introduced to
describe rare events, such as earthquakes or stock market crashes,
which act in one direction.  For example, earthquakes destroy property
but do not create it.  In the limit of large $K$, bosonic quasicondensates
have small
fluctuations and exhibit good interference patterns in most cases. But
occasionally there are strong fluctuations which lead to an appreciable
decrease of the contrast.
This gives rise to a strong asymmetry of the resulting distribution
function.
\begin{figure}[ht]
\includegraphics[width=8.0cm]{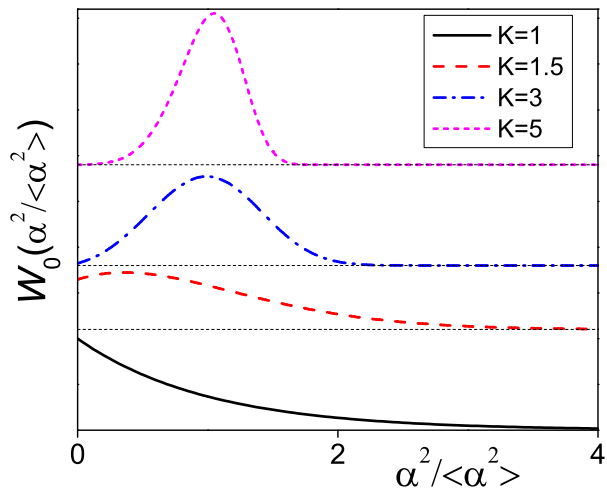}
\caption{Evolution of the distribution function
$W(\a)$ for different values of $K$.
At larger values of $K$ the function $W$ tends to the
delta-function (see the text).
}
\label{figK}
\end{figure}
\begin{figure}[ht]
\includegraphics[width=8.0cm]{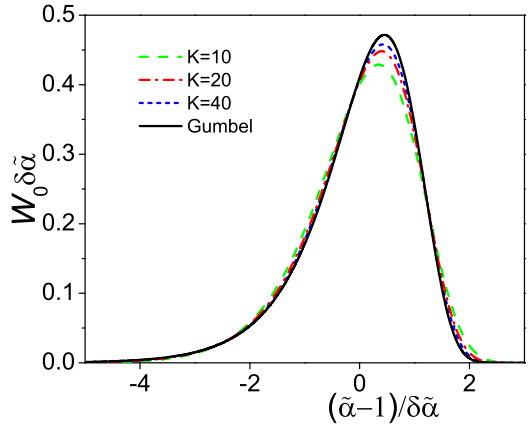}
\caption{Limit of large
$K$. Scaled distribution function $\delta\tilde\alpha
W\left((\tilde\a-1)/\delta\tilde\alpha\right)$, where
$\tilde{\alpha}=\a/\av{\a}=|A|^2/\av{|A|^2}$ and
$\delta\tilde\alpha$ is the width of the distribution.
The function $W$ is multiplied by $\delta\tilde\alpha$ to
preserve the total probability, which must be equal to unity. The
dashed and dotted lines correspond to different values of $K$. The
solid line corresponds to the conjectured Gumbel distribution.}
\label{figK1}
\end{figure}

The construction of \cite{BLZ1-3} is based on the representation
space of the Virasoro algebra for the set of central charges
satisfying $c=1-6(\sqrt{2K}-1/\sqrt{2K}))^{2}$ and the highest
weight $\Delta=(c-1)/24$ which gives $c\leq -2$ for $K\geq1$. For
all $K>1$ the central charge of the Virasoro algebra is negative.
Theories with negative central charges appear in different contexts
in statistical mechanics, stochastic growth models, 2D quantum
gravity, models of 2D turbulence and even high-energy QCD. The
experimentally measured distribution function $W$ can be inverted
(using the Bessel functions completeness relation) to obtain the
$Q$-operator which in turn can be used to reconstruct the transfer
matrices of the above mentioned  models with negative $c$.
Therefore, the interference of condensates provides a possible way to
explore the interesting physics of various models ranging from
statistical to high energy physics.

To summarize, in this paper we discussed interference experiments
between two independent one-dimensional quasi-condensates. We showed
that quantum phase fluctuations act to suppress the average
interference contrast and to induce fluctuations in this quantity from
one experimental run to the next. The average interference contrast depends
on the two point correlation. It therefore scales as a non trivial
power of the imaging length related to the power law decay of the correlations.
The distribution function characterizing the shot to shot fluctuations
of the contrast contains information on high order correlation functions.
We computed
the distribution function of the amplitude of interference fringes
using the relation of this problem to the partition function of the
logarithmic Coulomb gas on a circle and to the properties of $Q$
operators of conformal field theories with negative central
charges. We showed that the distribution function of fringe amplitudes
is related to non-trivial physical properties of a variety of
interesting statistical and field-theoretical models.

{\bf Acknowledgments} We are grateful to P.~Fendley, V. M.
Galitski, M.~Greiner, Z.~Hadzibabic, M.~Lukin, S. L. Lukyanov,
M.~Oberthaller, M.~Oshikawa, M. Pletyukhov, J.~Schmiedmayer,
V.~Vuletic, D.~Weiss, K.~Yung and A. B. Zamolodchikov for useful
discussions. This work was partially supported by the NSF grant
DMR-0132874. V.G. is supported by Swiss National Science Foundation.


\begin{thebibliography}{9}
\bibitem{note} We point out that Eq.(\ref{z2n}) corresponds to
periodic boundary condition. This may give a small quantitative
change but will not affect the qualitative picture of the evolution
of the distribution function with $K$.
\end{thebibliography}
\end{document}